\documentclass[apjl]{emulateapj}
\pdfoutput=1

\usepackage[fleqn]{amsmath} 
\usepackage{amsfonts} 
\usepackage{amstext}
\usepackage{amssymb} 
\usepackage[colorlinks=true, citecolor=blue, linkcolor=blue, breaklinks=true]{hyperref}

\newcommand{\hi } {{\rm H}\,{\small\rm I} \,}

\begin{document}

\title{The small scatter of the baryonic Tully-Fisher relation}

\author{Federico Lelli\altaffilmark{1}, Stacy S. McGaugh\altaffilmark{1}, James M. Schombert\altaffilmark{2}}
\altaffiltext{1}{Department of Astronomy, Case Western Reserve University, Cleveland, OH 44106, USA, \email{federico.lelli@case.edu}}
\altaffiltext{2}{Department of Physics, University of Oregon, Eugene, OR 97403, USA}


\begin{abstract}
In a $\Lambda$CDM cosmology, the baryonic Tully-Fisher relation (BTFR) is expected to show significant intrinsic scatter resulting from the mass-concentration relation of dark matter halos and the baryonic-to-halo mass ratio. We study the BTFR using a sample of 118 disc galaxies (spirals and irregulars) with data of the highest quality: extended \hi rotation curves (tracing the outer velocity) and Spitzer photometry at 3.6~$\mu$m (tracing the stellar mass). Assuming that the stellar mass-to-light ratio ($\Upsilon_{*}$) is nearly constant at 3.6~$\mu$m, we find that the scatter, slope, and normalization of the BTFR systematically vary with the adopted $\Upsilon_{*}$. The observed scatter is minimized for $\Upsilon_{*}\gtrsim0.5$ $M_{\odot}/L_{\odot}$, corresponding to nearly maximal discs in high-surface-brightness galaxies and BTFR slopes close to $\sim$4. For any reasonable value of $\Upsilon_{*}$, the intrinsic scatter is $\sim$0.1 dex, below general $\Lambda$CDM expectations. The residuals show no correlations with galaxy structural parameters (radius or surface brightness), contrary to the predictions from some semi-analytic models of galaxy formation. These are fundamental issues for $\Lambda$CDM cosmology.
\end{abstract}

\keywords{galaxies: kinematics and dynamics --- galaxies: formation --- galaxies: evolution --- galaxies: spiral --- galaxies: irregular --- dark matter}

\section{Introduction}\label{sec:Intro}

The baryonic Tully-Fisher relation (BTFR) links the rotation velocity of a galaxy to its total baryonic mass ($M_{\rm b}$) and extends for 6 decades in $M_{\rm b}$ \citep{McGaugh2012}. In a $\Lambda$ cold dark matter ($\Lambda$CDM) cosmology, the BTFR must emerge from the complex process of galaxy formation, hence it is expected to show significant intrinsic scatter. Using a semi-analytic galaxy-formation model, \citet{Dutton2012} predicts a \textit{minimum} intrinsic scatter of $\sim$0.15 dex along the BTFR \citep[see also][]{DiCintio2015}. The majority of this scatter comes from the mass-concentration relation of dark matter (DM) halos, which is largely independent of baryonic processes and well-constrained by cosmological DM-only simulations \citep[][]{Bullock2001}. Hence the BTFR scatter provides a key test for $\Lambda$CDM.

The BTFR has been extensively studied using \hi observations from radio interferometers \citep[e.g.,][]{Verheijen2001b, Noordermeer2007b} and single-dish telescopes \citep[e.g.,][]{Gurovich2010, Zaritsky2014}. Radio interferometers can spatially resolve the \hi kinematics in nearby galaxies, providing high-quality rotation curves (RCs). Despite being limited to small galaxy samples, the study of RCs has provided key insights on the BTFR. Primarily, the BTFR scatter is minimized using the velocity along the flat part ($V_{\rm f}$), which extends well beyond the optical galaxy \citep{Verheijen2001b, Noordermeer2007b}. Presumably, $V_{\rm f}$ is set by the DM halo and closely relates to its virial velocity.

Single-dish \hi surveys provide large samples for BTFR studies (thousands of objects), which are most useful to estimate galaxy distances and investigate local flows \citep{Tully2013}. When interpreted in a galaxy formation context, however, they are prone to systematic effects because rotation velocities from \hi line-widths do not necessarily correspond to $V_{\rm f}$ \citep{Verheijen2001b}. Studies using \hi line-widths generally reports shallower BTFR slopes ($\sim$3) than those using $V_{\rm f}$ ($\sim$4), leading to drastically different interpretations \citep{Gurovich2010, McGaugh2012}. Studies based on H$\alpha$ \citep{Pizagno2007} or CO \citep{Ho2007} observations may present similar issues, since H$\alpha$ and CO discs are typically less extended than \hi discs and their maximum velocities may not be tracing $V_{\rm f}$.

In this Letter, we investigate the BTFR using a sample of 118 galaxies with data of the highest quality: (i) extended \hi rotation curves providing precise measurements of $V_{\rm f}$, and (ii) Spitzer surface photometry at 3.6~$\mu$m providing the optimal tracer of the stellar mass.

\section{Data Analysis}\label{sec:Data}

\subsection{Galaxy Sample}\label{sec:Sample}

This work is based on the SPARC (Spitzer Photometry \& Accurate Rotation Curves) dataset, which will be presented in detail in Lelli et al. (in prep.). In short, we collected more than 200 high-quality \hi RCs of disc galaxies from previous compilations, large surveys, and individual studies. The major sources are \citet[][52 objects]{Swaters2009}, \citet[][30 objects]{Sanders1998}, \citet[][24 objects]{deBlok1997}, \citet[][22 objects]{Sanders1996}, and \citet[][17 objects]{Noordermeer2007}. The RCs were derived using similar techniques: fitting a tilted ring model to the \hi velocity field \citep[][]{Begeman1987} and/or using position-velocity diagrams along the disc major axis \citep{deBlok1997}.

Subsequently, we searched the Spitzer archive for 3.6$\mu$m images of these galaxies. We found 173 objects with useful [3.6] data. We derived surface brightness profiles and asymptotic magnitudes using the Archangel software \citep{Schombert2011}, following the same procedures as \citet{Schombert2014}. The surface brightness profiles will be presented elsewhere; here we simply use asymptotic magnitudes at [3.6] to estimate stellar masses ($M_{*}$).

For the sake of this study, we exclude starburst dwarf galaxies (8 objects from \citealt{Lelli2014a} and Holmberg~II from \citealt{Swaters2009}) because they have complex \hi kinematics and are likely involved in recent interactions \citep{Lelli2014b}. This reduces our starting sample to 164 objects.

\subsection{Rotation Velocity}\label{sec:Vrot}

Empirically, the velocity along the flat part of the RC minimizes the scatter in the BTFR \citep{Verheijen2001b}. The choice of velocities at some photometric radius, like 2.2 $R_{\rm d}$ \citep[the disc scale length,][]{Dutton2007} or $R_{80}$ \citep[the radius encompassing 80$\%$ of the i-band light,][]{Pizagno2007}, lead to relations with larger scatter. This can be simply understood. For some low-mass, gas-dominated galaxies, these radii may occur along the rising part of the RC \citep{Swaters2009}. For high-mass, bulge-dominated galaxies, they may occur along the declining part of the RCs \citep{Noordermeer2007b}. These photometric radii do not necessarily imply a consistent measurement of the rotation velocity due to the different distribution of baryons within galaxies. Conversely, $V_{\rm f}$ appears insensitive to the detailed baryonic distribution, being measured at large enough radii to encompass the bulk of the baryonic mass. $V_{\rm f}$ represents our best proxy for the halo virial velocity.

We estimate $V_{\rm f}$ using a simple automated algorithm. We start by calculating the mean of the two outermost points of the RC:
\begin{equation}
\overline{V} = \dfrac{1}{2}(V_{N} + V_{N-1}),
\end{equation}
and require that
\begin{equation}
\dfrac{|V_{N-2} - \overline{V}|}{\overline{V}} \leq 0.05. 
\end{equation}
If the condition is fulfilled, the algorithm includes $V_{N-2}$ in the estimate of $\overline{V}$ and iterates to the next velocity point. When the condition is falsified, the algorithm returns $V_{\rm f} = \overline{V}$. This algorithm returns similar values of $V_{\rm f}$ as previously employed techniques based on the logarithmic slope $\Delta \log(V)/\Delta \log(R)$ \citep{Stark2009}, but is more stable against small radial variations in the RC. Galaxies that do not satifsy the condition at the first iteration are rejected, hence we only consider RCs that are flat within $\sim$5$\%$ over at least three velocity-points. This excludes 36 galaxies with rising RCs ($\sim$20$\%$), reducing the sample to 129 objects.

The error on $V_{\rm f}$ is estimated as
\begin{equation}
 \delta_{V_{\rm f}} = \sqrt{\dfrac{1}{N} \sum^{N}_{i}\delta_{V_{i}}^{2} + \delta_{\overline{V}}^{2} + \left(\dfrac{V_{\rm f}}{\tan(i)}\delta_{i}\right)^{2}},
\end{equation}
where (i) $\delta_{V_{i}}$ is the random error on each velocity-point along the flat part of the RC, quantifying non-circular motions and kinematic asymmetries between the two sides of the disc; (ii) $\delta_{\overline{V}}$ is the dispersion around $\overline{V}$, quantifying the degree of flatness of the RC; and (iii) $\delta_{i}$ is the error on the outer disc inclination $i$. High-quality \hi velocity fields can be used to obtain kinematic estimates of $i$ and trace possible warps \citep[e.g.,][]{Battaglia2006}. This is another advantage of interferometric surveys over single-dish ones, since the latter must rely on photometric inclinations that depend on the assumed stellar-disc thickness and may be inappropriate for warped \hi discs. Inclination corrections become very large for face-on discs due to the $\sin(i)$ dependence, hence we exclude galaxies with $i<30^{\circ}$ reducing the sample to 118 galaxies. Clearly, this does not introduce any selection effects since galaxy discs are randomly oriented across the sky, but decreases the mean error on $V_{\rm f}$ to $\sim$0.03 dex.

\subsection{Baryonic Masses}\label{sec:Mbar}

We estimate the baryonic mass as
\begin{equation}
M_{\rm b}= M_{\rm g} + \Upsilon_{*} L_{[3.6]},
\end{equation}
where $M_{\rm g}$ is the gas mass, $L_{[3.6]}$ is the [3.6] luminosity \citep[adopting a solar magnitude of 3.24,][]{Oh2008}, and $\Upsilon_{*}$ is the stellar mass-to-light ratio. Several studies suggest that $\Upsilon_{*}$ is almost constant in the near infrared ([3.6] or $K$-band) over a range of galaxy types and masses. A small scatter of $\sim$0.1 dex is consistently found using different approaches: stellar population synthesis models \citep{McGaugh2014, Meidt2014}, resolved stellar populations \citep{Eskew2012}, and the vertical velocity dispersion of galaxy discs \citep{Martinsson2013}. Large inconsistencies persist in the overall normalization \citep{McGaugh2015}. In this Letter, we assume that $\Upsilon_{*}$ is constant among galaxies and systematically explore the BTFR for different $\Upsilon_{*}$.

In disc galaxies $M_{\rm g}$ is typically dominated by atomic gas (probed by \hi observations). The molecular gas content can be estimated from CO observations assuming a CO-to-H$_2$ conversion factor, which can vary from galaxy to galaxy depending on metallicity or other properties. CO emission is often undetected in low-mass, metal-poor galaxies \citep{Schruba2012}. Luckily, molecules generally are a minor dynamical component, contributing less than 10$\%$ to $M_{\rm b}$ \citep{McGaugh2015}. Similarly, warm/hot gas is negligible in the disc \citep{McGaugh2012}. Hence we assume $M_{\rm g} = 1.33 M_{\hi}$, where the factor 1.33 takes the contribution of Helium into account. We note that any contribution from ``dark gas'' (molecular and/or ionized) is implicitly included in $\Upsilon_{*}$ as long as this scales with $M_{*}$.

\begin{figure*}[t!]
\centering
\includegraphics[width=0.93\textwidth]{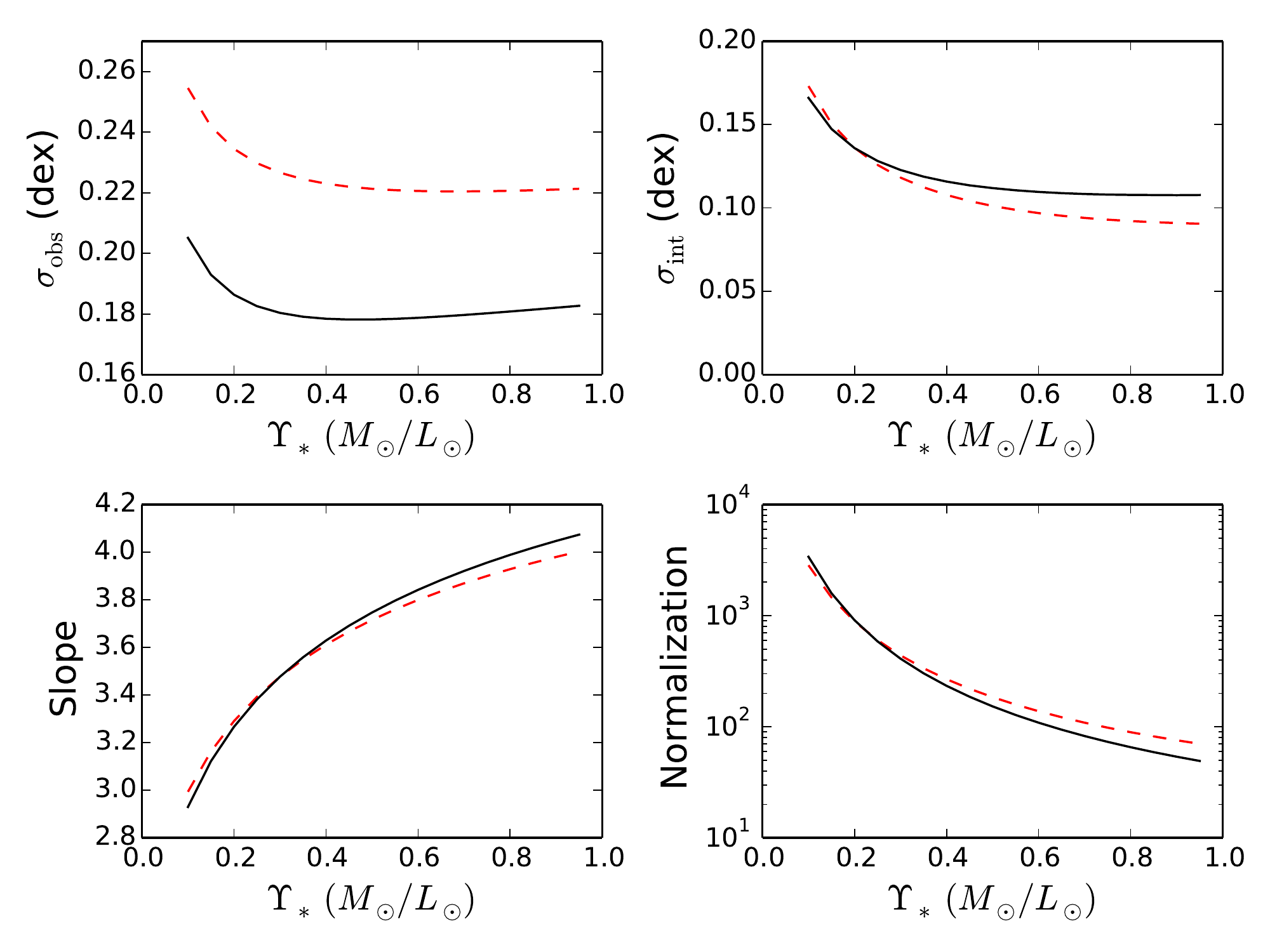}
\caption{The properties of the BTFR as a function of $\Upsilon_{*}$: observed scatter (top-left), intrinsic scatter (top-right), slope (bottom-left), and normalization (bottom-right). Dashed and solid lines show results for the total and accurate-distance samples, respectively.}
\label{fig:ML}
\end{figure*}
The error on $M_{\rm b}$ is estimated as
\begin{equation}
 \delta_{M_{\rm b}} = \sqrt{ \delta_{M_{\rm g}}^{2} + \bigg(\Upsilon_{*}\delta_{\rm L}\bigg)^{2} + \bigg(L \, \delta_{\Upsilon_{*}}\bigg)^{2} + \bigg(2 M_{\rm b}\dfrac{\delta_{D}}{D}\bigg)^{2}},
\end{equation}
where $\delta_{M_{\rm g}}$ and $\delta_{\rm L}$ are, respectively, the errors on $M_{\rm g}$ and $L_{[3.6]}$ due to uncertainties on total fluxes (generally smaller than $\sim$10$\%$). We assume $\delta_{\Upsilon_{*}}=0.11$~dex, as suggested by stellar population models \citep{Meidt2014, McGaugh2014}. Galaxy distances ($D$) and corresponding errors ($\delta_{D}$) deserve special attention. The 118 galaxies with accurate values of $V_{\rm f}$ have three different types of distance estimates (in order of preference):
\begin{enumerate}
 \item 32 objects have accurate distances from the tip of the red giant branch (26), Cepheids (3), or supernovae (3). These are generally on the same zero-point scale and have errors ranging from $\sim$5$\%$ to $\sim$10$\%$ \citep[see][]{Tully2013}. The majority of these distances (24/32) are drawn from the Extragalactic Distance Database \citep{Jacobs2009}.
 \item 26 objects are in the Ursa Major cluster, which has an average distance of $18 \pm 0.9$ Mpc \citep{Sorce2013}. For individual galaxies, one should also consider the cluster depth ($\sim$2.3~Mpc, \citealt{Verheijen2001b}), hence we adopt $\delta_{D} = \sqrt{2.3^2 + 0.9^2} = 2.5$~Mpc, giving an error of $\sim$14$\%$. We consider two galaxies from \citet{Verheijen2001b} as background/foreground objects: NGC~3992, having $D\simeq24$ Mpc from a Type-Ia supernova \citep{Parodi2000}, and UGC~6446, laying near the cluster boundary in both space and velocity ($D\simeq12$ Mpc from the Hubble flow).
  \item 60 objects have Hubble-flow distances corrected for Virgocentric infall (as given by NED\footnote{The NASA/IPAC Extragalactic Database (NED) is operated by the Jet Propulsion Laboratory, California Institute of Technology, under contract with the National Aeronautics and Space Administration.} assuming $H_{0} = 73$ km~s$^{-1}$~Mpc$^{-1}$). These are very uncertain for nearby galaxies where peculiar velocities may constitute a large fraction of the systemic velocities, but become more accurate for distant objects. Considering that peculiar velocities may be as high as $\sim$500~km~s$^{-1}$ and $H_{0}$ has an uncertainty of $\sim$7$\%$, we adopt the following errors: 30$\%$ for $D \leq 20$~Mpc; 25$\%$ for $20 < D \leq 40$~Mpc; 20$\%$ for $40 < D \leq 60$~Mpc; 15$\%$ for $60 < D \leq 80$~Mpc; and 10$\%$ for $D > 80$~Mpc. The most distant galaxy in our sample is UGC~1455 at $D\simeq130$~Mpc.
\end{enumerate}
We perform separate analyses for the total sample (mean $\delta_{M_{\rm b}}\simeq0.2$ dex) and for galaxies with accurate distances (58 objects in groups 1 and 2; mean $\delta_{M_{\rm b}}\simeq0.1$ dex).

\begin{figure*}
\centering
\includegraphics[width=\textwidth]{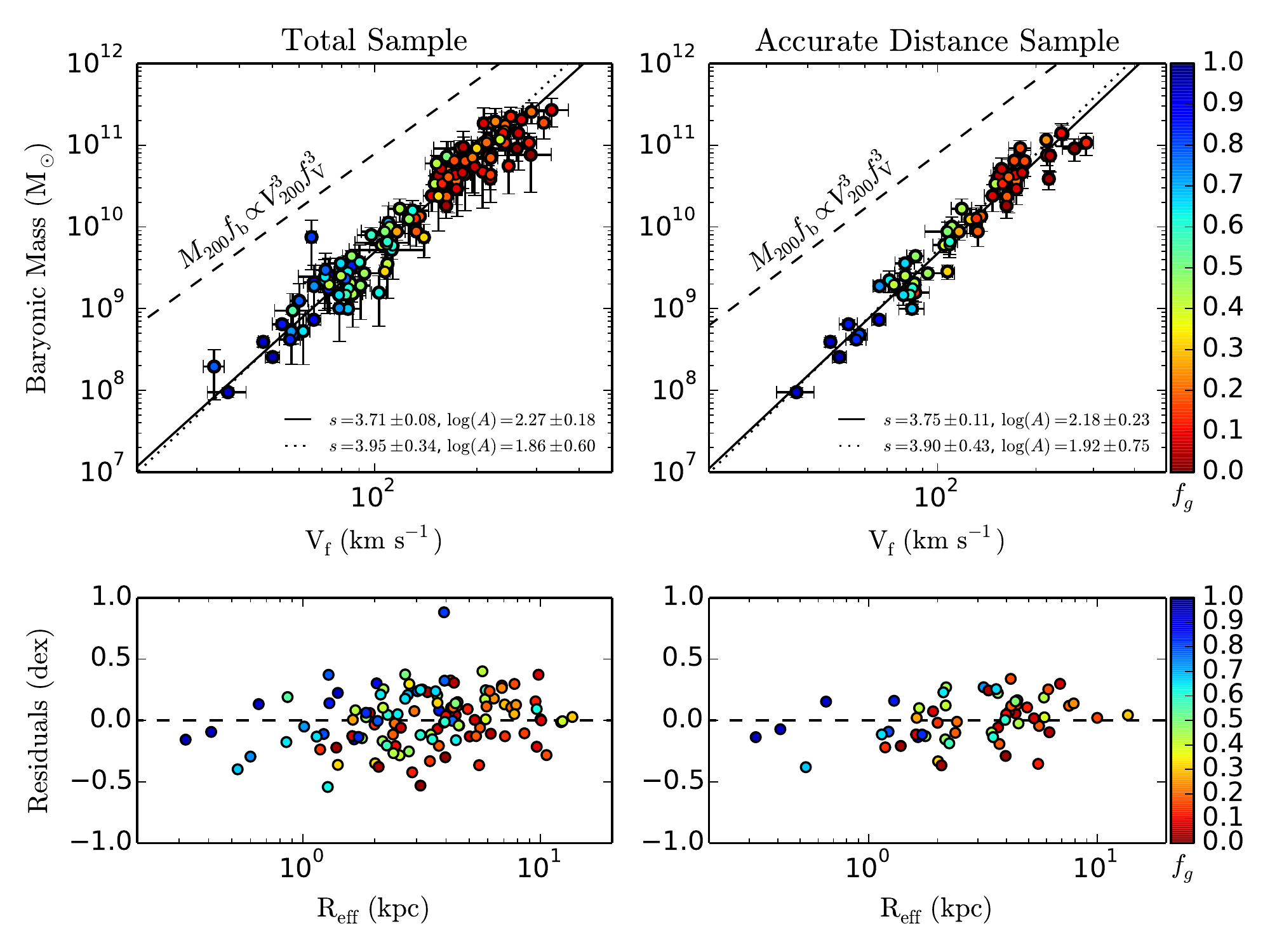}
\caption{\textit{Top panels}: BTFR adopting $\Upsilon_{*}=0.5$~$M_{\odot}/L_{\odot}$. Galaxies are color-coded by $f_{\rm g} = M_{\rm g}/M_{\rm b}$. Solid lines show error-weighted fits. Dotted lines show fits weighted by $f_{\rm g}^{2}$, increasing the importance of gas-dominated galaxies. The dashed line shows the $\Lambda$CDM initial condition with $f_{\rm V}=1$ and $f_{\rm b} = 0.17$ (the cosmic value). \textit{Bottom panels}: residuals from the error-weighted fits versus the galaxy effective radius. The outlier is UGC~7125, which has an unusually high correction for Virgocentric infall and lies near the region where the infall solution is triple-valued. If we consider only the correction for Local Group motion, UGC~7125 lies on the BTFR within the scatter.}
\label{fig:BTFR}
\end{figure*}

\section{Results}

We systematically explore the BTFR for different values of $\Upsilon_{*}$ ranging from 0.1 to 1.0 $M_{\odot}/L_{\odot}$ in steps of 0.05. Values higher than 1 are ruled out by maximum-disc models of high surface brightness (HSB) galaxies (Lelli et al. in prep.). For each $\Upsilon_{*}$, we fit a straight line:
\begin{equation}
\log(M_{\rm b}) = s \log(V_{\rm f}) + \log(A).
\end{equation}
We use the LTS\_LINEFIT algorithm \citep{Cappellari2013}, which performs a least-square minimization considering errors in both variables and allowing for intrinsic scatter. Figure~\ref{fig:ML} shows the observed scatter ($\sigma_{\rm obs}$), intrinsic scatter ($\sigma_{\rm int}$), slope ($s$), and normalization ($A$) as a function of $\Upsilon_{*}$.

\subsection{Observed and Intrinsic Scatter}

Figure~\ref{fig:ML} (top-left) shows that $\sigma_{\rm obs}$ decreases with $\Upsilon_{*}$ and reaches a plateau at $\Upsilon_{*}\gtrsim0.5 M_{\odot}/L_{\odot}$. This plateau actually is a broad minimum that becomes evident by extending the $\Upsilon_{*}$-range up to unphysical values of $\sim$10 $M_{\odot}/L_{\odot}$ (not shown). The observed scatter is systematically lower for the accurate-distance sample, indicating that a large portion of $\sigma_{\rm obs}$ in the full sample is driven by distance uncertainties.

Figure\ref{fig:ML} (top-right) shows that $\sigma_{\rm int}$ is below 0.15 dex for any realistic value of $\Upsilon_{*}$. The similar intrinsic scatter between the two samples suggests that our errors on Hubble-flow distances are realistic. For a fiducial value of $\Upsilon_{*} = 0.5$ $M_{\odot}/L_{\odot}$, we find $\sigma_{\rm int} = 0.10\pm0.02$ for the full sample and $\sigma_{\rm int} = 0.11 \pm 0.03$ for the accurate-distance sample. As we discuss in Sect.~\ref{sec:disc}, this represents a challenge for the $\Lambda$CDM cosmological model. 

\subsection{Slope, Normalization, and Residuals}

Figure\ref{fig:ML} (bottom panels) shows that the BTFR slope (normalization) monotonically increases (decreases) with $\Upsilon_{*}$. This is due to the systematic variation of the gas fraction ($f_{\rm g} = M_{\rm g}/M_{\rm b}$) with $V_{\rm f}$. Figure~\ref{fig:BTFR} (top panels) shows the BTFR for $\Upsilon_{*} = 0.5$ $M_{\odot}/L_{\odot}$, colour-coding each galaxy by $f_{\rm g}$. Low-mass galaxies tend to be gas-dominated ($f_{\rm g} \gtrsim 0.5$) and their location on the BTFR does not strongly depend on the assumed $\Upsilon_{*}$ \citep{Stark2009, McGaugh2012}. Conversely, high-mass galaxies are star-dominated and their location on the BTFR strongly depends on $\Upsilon_{*}$. By decreasing $\Upsilon_{*}$, $M_{\rm b}$ decreases more significantly for high-mass galaxies than for low-mass ones, hence the slope decreases and the normalization increases.

For any $\Upsilon_{*}$, we find no correlation between BTFR residuals and galaxy effective radius: the Pearson's, Spearman's, and Kendall's coefficients are consistently between $\pm$0.4. Figure~\ref{fig:BTFR} (bottom panels) shows the case of $\Upsilon_{*}=0.5$ $M_{\odot}/L_{\odot}$. Similarly, we find no trend with effective surface brightness. We have also fitted exponentials to the outer parts of the luminosity profiles and find no trend between residuals and central surface brightness or scale length. These results differ from those of \citet{Zaritsky2014} due to the use of $V_{\rm f}$ instead of \hi line-widths \citep[see also][]{Verheijen2001b}. The lack of any trend between BTFR residuals and galaxy structural parameters is an issue for galaxy formation models, which generally predict such correlations \citep{Dutton2007, Desmond2015}.

Gas-dominated galaxies provide an absolute calibration of the BTFR. \citet{McGaugh2012} used 45 galaxies that were selected for having $f_{\rm g} > 0.5$ (assuming $\Upsilon_{*}$ based on stellar population models). Here we use a slightly different approach: we fit the BTFR weighting each point by $1/f^{2}_{\rm g}$. In this way, galaxies with $f_{\rm g}=0.5$ are 25 times more important in the fit than galaxies with $f_{\rm g} = 0.1$. Our results are shown in Figure~\ref{fig:BTFR} for $\Upsilon_{*}=0.5$ $M_{\odot}/L_{\odot}$, using both standard error-weighted fits and $f_{\rm g}^2$-weighted fits. The latter hints at a small, systematic increase of $\Upsilon_{*}$ with $1/f_{\rm g}$ or $V_{\rm f}$ (by $\sim$0.1 dex) in order to have symmetric residuals.

\section{Discussion \& Conclusions}\label{sec:disc}

In this Letter we study the BTFR using 118 galaxies with data of the highest quality: extended \hi RCs tracing $V_{\rm f}$ and Spitzer photometry at 3.6~$\mu$m tracing $M_{*}$. We systematically explore the BTFR for different values of $\Upsilon_{*}$ and find the following results:
\begin{enumerate}
 \item The observed scatter reaches a broad minimum for $\Upsilon_{*} \gtrsim 0.5 M_{\odot}/L_{\odot}$, corresponding to nearly maximal stellar discs in HSB galaxies;
 \item For any reasonable value of $\Upsilon_{*}$, the intrinsic scatter is below the \textit{minimum} value expected in a $\Lambda$CDM cosmology \citep[0.15 dex,][]{Dutton2012};
 \item The residuals around the BTFR show no trend with galaxy size or surface brightness, providing a further challenge to galaxy formation models \citep{Desmond2015}.
\end{enumerate}

In a $\Lambda$CDM cosmology, the BTFR results from an underlying correlation between the ``virial'' mass ($M_{\rm 200}$) and ``virial'' velocity ($V_{200}$) of the DM halo \citep{Mo1998}. Following \citet{McGaugh2012}, we write 
\begin{equation}
M_{200} = (\sqrt{100}G \, H_{0})^{-1} V_{200}^3,
\end{equation}
where $G$ and $H_{0}$ are Newton's and Hubble's constants, respectively. By construction, this relation has no intrinsic scatter. To map $M_{200}$ and $V_{200}$ into the observed BTFR, we introduce the baryonic fraction $f_{\rm b} = M_{\rm b}/M_{\rm 200}$ and the factor $f_{\rm V} = V_{\rm f}/V_{200}$, hence 
\begin{equation}
M_{\rm b} \propto f_{\rm b} f_{\rm V}^{-3} V_{\rm f}^{3}.
\end{equation}
Since the observed BTFR has a slope higher than 3, we infer that $f_{\rm b}$ and/or $f_{\rm V}$ systematically vary with $M_{\rm b}$. The presumed $f_{\rm b}-M_{\rm b}$ and $f_{\rm V}-M_{\rm b}$ relations are set by the galaxy formation process and must induce scatter on the observed BTFR. The factor $f_{\rm b}$ is determined by baryonic processes like gas inflows, reionization, stellar feedback and gas outflows. The factor $f_{\rm V}$, instead, is determined both by ``primordial'' halo properties (mass-concentration relation) and subsequent baryonic physics (halo contraction due to baryonic infall or expansion due to stellar feedback).

The mass-concentration relation has a well-established scatter of 0.11 dex from N-body simulations \citep{Dutton2014}, driving scatter on $f_{\rm V}$. The $M_{*}-M_{200}$ relation, proxy of a more fundamental $M_{\rm b}-M_{200}$ relation, has an estimated scatter of $\sim$0.1-0.2 dex \citep{Moster2013, Zu2015}, driving scatter on $f_{\rm bar}$. Since these two sources of scatter act in perpendicular directions on the BTFR, we expect an intrinsic scatter of \textit{at least} $\sim$0.15-0.2 dex. For example, \citet{Dutton2012} uses a semi-analytic model of galaxy formation to reproduce the BTFR and predicts a minimum intrinsic scatter of 0.15 dex. In his model, most of the expected scatter ($\sim$73$\%$) comes from the mass-concentration relation, while the remaining fraction comes from variations in the halo spin parameter. Hence, a value of $\sim$0.15 is a \textit{lower limit} for $\Lambda$CDM cosmology: it is hard to imagine that stochastic baryonic processes would conspire to reduce the scatter expected from the basic structure of DM halos. Conversely, the diverse formation histories of galaxies should lead to significant variations in $f_{\rm b}$ at a given halo mass \citep{Eisenstein1996, McGaugh1998}, hence a scatter larger than 0.15 dex would be a more natural result.

We find that the BTFR intrinsic scatter is $\sim$0.11 dex, below $\Lambda$CDM expectations. For galaxies with accurate distances, the observed scatter is already small ($\sim$0.18 dex), thus there is little room for having overestimated the errors. Moreover, $\sigma_{\rm int}\simeq0.11$ dex is comparable to the mean error on $M_{\rm b}$ (driven by the assumed uncertainty on $\Upsilon_{*}$), hence it should be considered as an observational \textit{upper limit}.

One may wonder whether selection effects may artificially decrease the BTFR scatter. For example, \citet{Gurovich2010} argued that the exclusion of low-mass galaxies with rising RCs may introduce a bias towards objects where the \hi distribution is more extended with respect to the DM scale length. This is hardly the case. For low-mass bulgeless galaxies, RCs start to flatten at $\sim$2~$R_{\rm d}$ (independently of mass or luminosity) and reach a flat part beyond $\sim$3~$R_{\rm d}$ \citep{Swaters2009}. The presence or not of a flat part is mostly related to our ability to trace RCs out to large radii, where the \hi column densities are low and the sensitivity of the observations decreases. Hence, the exclusion of galaxies with rising RCs helps to \textit{avoid} observational bias. 

We stress that our sample spans a large range in mass ($10^{8}\lesssim M_{\rm b}/M_{\odot} \lesssim 10^{11}$), size ($0.3 \lesssim R_{\rm eff}/\rm{kpc} \lesssim 15$), surface brightness ($4 \lesssim \Sigma_{\rm eff}/L_{\odot} \, \rm{pc^{-2}} \lesssim 1500$), gas fraction ($0.01 \lesssim f_{\rm g} \lesssim 0.95$), and morphology (S0 to Im). We are only excluding merging and interacting galaxies, since they show disturbed, out-of-equilibrium \hi kinematics that would artificially inflate the scatter on the BTFR. Hence our results are representative for the general population of non-interacting galaxies in low-density environments (field, groups, and diffuse clusters like Ursa Major).

In conclusion, the BTFR is an open issue for the current cosmological model: the stochastic process of galaxy formation needs to reproduce a global relation with little (if any) intrinsic scatter and no dependence on galaxy structural parameters.

\acknowledgments


We thank Michele Cappellari for providing LTS\_LINEFIT. The work of FL and SSM is supported by a grant from the John Templeton Foundation.


\end{document}